\documentclass{ws-ijmpb}

\begin{document}

\markboth{M. A. da Silva Jr., R. M. Serra and L. C. C\'{e}leri}
{Observer invariance of the collapse postulate of quantum mechanics}

\catchline{}{}{}{}{}
%
%%%%%%%%%%%%%%%%%%%%%%%%%%%%%%%%%%%%%%%%%%%%%%%%%%%%%%%%%%%%%%%%%%%%

\title{Observer invariance of the collapse postulate of quantum mechanics}

\author{Milton A. da Silva Jr.}
\address{Centro de Ci\^{e}ncias Naturais e Humanas, Universidade Federal do ABC, Rua Santa
Ad\'{e}lia 166, 09210-170, Santo Andr\'{e}, S\~{a}o Paulo, Brazil}

\author{Roberto M. Serra}
\address{Centro de Ci\^{e}ncias Naturais e Humanas, Universidade Federal do ABC, Rua Santa
Ad\'{e}lia 166, 09210-170, Santo Andr\'{e}, S\~{a}o Paulo, Brazil}

\author{Lucas C. C\'{e}leri}
\address{Centro de Ci\^{e}ncias Naturais e Humanas, Universidade Federal do ABC, Rua Santa
Ad\'{e}lia 166, 09210-170, Santo Andr\'{e}, S\~{a}o Paulo, Brazil\\
lucas@chibebe.org}

\maketitle

%\begin{history}
%\received{Day Month Year}
%\revised{Day Month Year}
%\accepted{(Day Month Year)}
%\comby{(xxxxxxxxxx)}
%\end{history}

\begin{abstract}
We analyse the wave function \textit{collapse} as seem by two distinct observers (with identical detectors) in relative motion. Imposing that the measurement process demands information transfer from the system to the detectors, we note that although different observers will acquire different amount of information from their measurements due to correlations between spin and momentum variables, all of them will agree about the orthogonality of the outcomes, as defined by their own reference frame. So, in this sense, such a quantum mechanical postulate is observer invariant, however the \textit{effective efficiency of the measurement process} differs for each observer.
\end{abstract}


\section{Introduction} 
The first quarter of twentieth century witnessed the development of two theories that lie on the basis of modern physics and technology: Quantum mechanics and the theory of relativity. Quantum mechanics is based on some postulates that can be stated, in general, as \cite{Cohen,Bohr}: ($i$) the state of a quantum system is represented by a vector in its corresponding Hilbert space; ($ii$) quantum evolutions are unitary; ($iii$) immediate repetition of a measurement yields the same outcome; ($iv$) outcomes are restricted to an orthonormal set $\left\{  \left\vert a_{k}\right\rangle \right\}  $ of eigenstates of the measured observable (\emph{collapse postulate}); ($v$) if the system is in state $\left\vert \psi\right\rangle $ before the measurement, the probability of a given outcome $a_{k}$ (associated to the eigenvector $\left\vert a_{k}\right\rangle $) is $p_{k}=\left\vert \left\langle a_{k}|\psi \right\rangle \right\vert ^{2}$ (Born's rule). The first two postulates imply linearity (which is in the core of the celebrated \textit{no-cloning} theorem \cite{NonCloning}) while the remaining three postulates establish the connection between state vectors and measurements of physical quantities (referred as observables). On the other hand the special theory of Relativity is built upon two postulates \cite{Weinberg}: The first one is the fact that the speed of light is an universal constant and the second one state that the laws of physics have the same form in all inertial frames. Here we are interested in the consequences of this latter postulate (also known as Lorentz invariance) on the \emph{collapse} of the quantum mechanical wave function.

There is a long-standing debate over the apparent conflict between the \emph{collapse postulate} and the relativistic invariance (see, for example, \cite{Aharonov,Peres,PeresBook}). The problem arises due to the fact that quantum measurement processes affect the quantum state {}\textquotedblleft instantaneously\textquotedblright{}\ throughout the entire space and this is interpreted by some authors as a violation of Lorentz invariance. One may argue that the quantum state is not a physical entity and thus, need not to be invariant, but we can certainly ask ourselves about the dynamics of a quantum system when observed (measured) in different reference frames. We do not intend to enter in the discussion about the reality of the wave function, instead we are interested in the question of how different observers (in different inertial frames) perceive the wave-package collapse. Peres, Scudo and Terno \cite{Peres1} showed that the reduced spin density matrix of a particle is not invariant under a Lorentz transformation, only the entire density matrix is invariant. This has the consequence that different observers will notice different statistical distributions of clicks in their spin detectors. Studying the amount of information transferred from the system to the measurement apparatus, Zurek \cite{Zurek} argued that the \emph{collapse postulate} is a consequence of the linearity and the unitarity of the quantum theory.

Our aim in this article is to discuss the collapse postulate as seen by different observers in the framework of quantum information theory (QIT). We obtain that, although the statistical distribution of clicks are different for observers in distinct inertial frames, the fact that only orthogonal states can leave imprints on the measurement apparatus is invariant under Lorentz transformations. In other words, each observer will extract a different amount of information from the quantum system, but all observers will agree about the orthogonality of the outcomes. Such an issue could be crucial in the design of protocols for quantum communication involving different inertial frames. Besides the fundamental interest, such a discussion may\ also have practical importance due to new trends of implementing quantum information protocols in global scales~\cite{Zeilinger,Pan,Weinfurter,Zeilinger2,Zeilinger3}.

\section{Quantum states in different reference frames} 
For simplicity we will consider here a qubit encoded in a massive spin-$1/2$ particle initially prepared in a pure state. The normalized quantum state of such a qubit, in the momentum representation ($\mathbf{p}$), may be written as
\begin{equation}
|\Psi\rangle=\sum_{r}\int d\mathbf{p}\,a_{r}\left(  \mathbf{p}\right) |r,\mathbf{p}\rangle, \label{eq:quantum-state}
\end{equation}
where the amplitudes $a_{r}\left(  \mathbf{p}\right)  $ satisfy $\sum_{r}\int\left\vert a_{r}\left(  \mathbf{p}\right)  \right\vert ^{2}d\mathbf{p}=1$, $r=1,2$.

We are interested here in the spin reduced density matrix, which is obtained by tracing out the momentum variables from the complete state
\begin{equation}
\rho_{S}=\int d\mathbf{p}\psi(\mathbf{p})\psi(\mathbf{p})^{\dagger}=\frac{1}{2}\left(
\begin{array}
[c]{cc}
1+\gamma & \delta\\ 
\delta^{\ast} & 1-\gamma
\end{array}
\right), 
\label{reduced}
\end{equation}
where we have defined $\gamma\equiv\int d\mathbf{p}\left(  \left\vert a_{1}(\mathbf{p})\right\vert ^{2}-\left\vert a_{2}(\mathbf{p})\right\vert ^{2}\right)  $, $\delta\equiv2\int d\mathbf{p}a_{1}(\mathbf{p})a_{2}(\mathbf{p})^{\ast}$ \cite{Peres,Peres1} and $\psi(\mathbf{p})=\left\langle \mathbf{p}\left\vert \Psi\right.  \right\rangle $. The von Neumann entropy associated with this density operator is given by $S=-\sum_{i}\lambda_{i}\ln\lambda_{i}$, with $\lambda_{i}=\left(  1\pm\sqrt{\gamma^{2}+\delta^{2}}\right)  /2$ being the eigenvalues of $\rho_{S}$.

\begin{figure}[h]
\begin{center}
\includegraphics[scale=0.35]{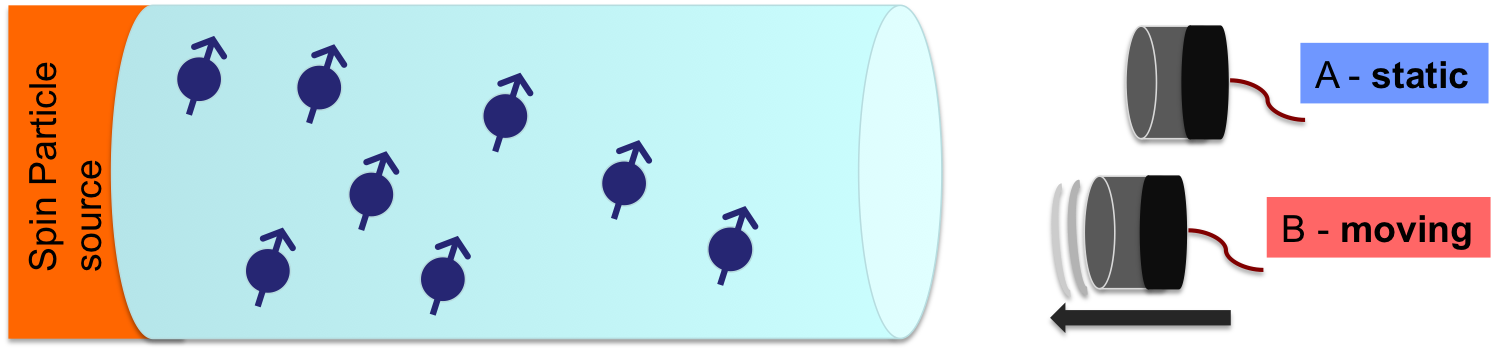}
\caption{(Colour Online) Schematic diagram of the considered scenario.}
\end{center}
\end{figure}

Let us consider two observers, one of them static in the qubit's source rest frame and a second one moving with constant velocity $v$ with respect to the qubit's source rest frame, as sketched in figure 1. The transformed state of the particle in the moving observer's frame is
\begin{equation}
|\Phi\rangle=\sum_{r}\int d\mathbf{p}\, b_{r}\left(  \mathbf{p}\right)|r,\mathbf{p}\rangle, 
\label{eq:transformed-quantum-state}
\end{equation}
with the following components \cite{Weinberg:1996,Halpern:1968}:
\begin{equation}
b_{r}\left(  \mathbf{p}\right)  =\left[  \frac{\left(  \Lambda^{-1}p\right)^{0}}{p^{0}}\right]  ^{1/2}\sum_{r^{\prime}}D_{r,r^{\prime}}\left[\Lambda,\Lambda^{-1}p\right]  a_{r^{\prime}}\left(  \Lambda^{-1}p\right),
\label{eq:spinor-component-transformation}
\end{equation}
where $p=\left(  p^{0},\mathbf{p}\right)  $ and $D_{r,r^{\prime}}$ are the elements of the Wigner rotation matrix $D\left[  \Lambda,p\right]=L^{-1}\left(  \Lambda p\right)  \Lambda L\left(  p\right)  $ \cite{Wigner}, for a general Lorentz transformation  $\Lambda$ and a pure boost $L$ \cite{Weinberg:1996,Halpern:1968}. For the case of a massive particle and a Lorentz transformation in the $x$-direction considered here, we have (adopting the Einstein summation rule)
\begin{align}
D\left[  \Lambda,p\right]   &  =\frac{\left(  p^{0}+m\right)  \sigma^{0}\cosh\left(  \theta/2\right)  }{\left\{  \left(  p^{0}+m\right)  \left[\left(  \Lambda p\right)  ^{0}+m\right]  \right\}  ^{1/2}}+\nonumber\\
&  +\frac{\left(  p^{x}\sigma^{0}+\text{i}\epsilon_{xij}p^{i}\sigma^{j}\right)  \sinh\left(  \theta/2\right)  }{\left\{  \left(  p^{0}+m\right)\left[  \left(  \Lambda p\right)  ^{0}+m\right]  \right\}  ^{1/2}},
\label{eq:Wigner-rotation-matrix}
\end{align}
where $\sigma^{0}$ and $\sigma^{i},\,\, i=x,y,z$ are the usual $2\times2$ identity and Pauli matrices, respectively, and we have defined $\theta \equiv-\tanh^{-1}v$.

In order to evaluate how these transformations lead us to the fact that observers in different Lorentz frames will perceive different amounts of information, let us consider that each of the two observers has identical apparatus that detects the spin of particles emitted by a source $\mathbb{S}$. Let us suppose that these qubits are prepared in the positive eigenstates of $\sigma^{x}$ such that $a_{1}\left(  \mathbf{p}\right) =a_{2}\left(  \mathbf{p}\right)  =a\left(  \mathbf{p}\right)  /\sqrt{2}$. By taking the partial trace of this state over the momenta we obtain the reduced density matrix for the spin variables, which can be measured by the apparatuses. For the static observer this is a pure state (its entropy is zero). The description of such a state by the observer that moves with velocity $v$ with respect to the source $\mathbb{S}$ is given by a statistical mixture $\rho_{S}$ with the following amplitudes
\begin{subequations}
\begin{align}
b_{1}\left(  \mathbf{p}\right)   &  =\frac{K}{\sqrt{2}}\left(  \frac{q^{0}}{p^{0}}\right)  ^{1/2}\left[  \left(  q^{0}+m\right)  \cosh\left(\frac{\theta}{2}\right)  \right.  +\nonumber\\
&  +\left.  \left(  q^{x}-q^{z}+\text{i}q^{y}\right)  \sinh\left(\frac{\theta}{2}\right)  \right]  a\left(  \mathbf{p}\right),
\label{eq:transformed-amplitudes-a}\\
b_{2}\left(  \mathbf{p}\right)   &  =\frac{K}{\sqrt{2}}\left(  \frac{q^{0}}{p^{0}}\right)  ^{1/2}\left[  \left(  q^{0}+m\right)  \cosh\left(\frac{\theta}{2}\right)  \right.  +\nonumber\\
&  +\left.  \left(  q^{x}+q^{z}-\text{i}q^{y}\right)  \sinh\left(\frac{\theta}{2}\right)  \right]  a\left(  \mathbf{p}\right),
\label{eq:transformed-amplitudes}
\end{align}
where $K=\left[  \left(  q^{0}+m\right)  \left(  p^{0}+m\right)  \right]^{-1/2}$ with $q^{\mu}=\left(\Lambda^{-1}p\right)  ^{\mu}$ ($\mu=0,x,y,z$). For this observer, the spin reduced density matrix is given by Eq. (\ref{reduced}) replacing the coefficients $a_{i}$ by $b_{i}$.

Assuming that, in the Lorentz frame of the qubit's source, the momentum has a Gaussian distribution of the form
\end{subequations}
\begin{equation}
a\left(  \mathbf{p}\right)  =\pi^{-3/4}\,\,\xi^{-3/2}\exp\left(-\frac{\mathbf{p}^{2}}{2\xi^{2}}\right),
\label{momentum-dist}
\end{equation}
and that this distribution is very narrow, i.e., $\xi/m\ll1$, we are able to compute the entropy of the reduced spin system as
\begin{equation}
S=-\frac{1}{2}\left[  \left(  1+\delta\right)  \ln\left(  1+\delta\right) +\left(  1-\delta\right)  \ln\left(  1-\delta\right)  \right]  +\ln2,
\label{entropy}
\end{equation}
with
\begin{equation}
\delta=1-\frac{\xi^{2}}{8m^{2}}\tanh^{2}\left(  \frac{\theta}{2}\right).
\label{eq:delta}
\end{equation}
For a static observer where $v=0$ (which implies $\theta=0$), we obtain the expected result for an uncorrelated state, i.e., $\delta=1$ which leads us to $S=0$. We also note that, in the case of a system in a momentum eigenstate, i.e., $\xi=0$, the motion of the detector does not influence the information gain in the measurement processes. This is due to the fact that, in this case, the Wigner rotation does not correlate the spin and momentum degrees of freedom \cite{Peres,Peres1,Landulfo,Kim}.

From Eqs. (\ref{entropy}) and (\ref{eq:delta}), we straightforwardly conclude that for not null velocities $v$, the entropy $S$ is greater than zero. This implies that the two observers, although equipped with identical detectors, will obtain a different probability distribution when measuring the spin component of the particle. While the observer in the qubit rest frame of $\mathbb{S}$ perceives a pure state with zero entropy, the moving one will perceive a mixed state, with non-zero entropy. In the framework of QIT, we could say that the amount of information that both observers have access can be quite different depending on the choice of the Lorentz frame and on the observable which is measured. This is a consequence of the correlations between spin and momentum variables introduced by the Wigner rotation, so when we take the partial trace over the momentum some information about the spin is lost.

The fact that the outcomes of a measurement are not invariant under a Lorentz transformation (i.e. the click distribution is not observer invariant), leads us to the natural question of how it affects the wave function collapse, since both the outcome of a measurement and the wave function collapse are closely related. In what follows, we point out that the wave function collapse postulate, interpreted in terms of QIT \cite{Zurek}, is independent of the Lorentz frame adopted by the observer. More specifically, we observe that the fact that only orthogonal states (as defined by each observer) can leave imprints on the measurement apparatus is Lorentz invariant.

\section{Orthogonality of the Outcomes} 
A measurement process can be regarded as a transfer of information from the system $\mathcal{S}$ to the measurement apparatus $\mathcal{A}$ in such way that only orthogonal states can leave imprints on $\mathcal{A}$ \cite{Zurek}. This kind of breaking on the unitary evolution provides a framework for the wave function collapse. Following the reasoning presented in Ref. \cite{Zurek}, we consider a two-level system $\mathcal{S}$, whose state can be written as a superposition of a pair of linearly independent kets $|\psi_{\mathcal{S}}\rangle=\alpha\left\vert \nwarrow\right\rangle +\beta\left\vert \nearrow\right\rangle $ (not necessarily orthogonal) interacting with a measurement apparatus $\mathcal{A}$ that starts in a blank state $|A_{blank}\rangle$.

The information transfer from $\mathcal{S}$ to $\mathcal{A}$ after the measurement process can be represented through the relation $\left(\alpha\left\vert \nwarrow\right\rangle +\beta\left\vert \nearrow\right\rangle\right)  \left\vert A_{blank}\right\rangle \Rightarrow\left\vert\Phi_{\mathcal{SA}}\right\rangle =\alpha\left\vert \nwarrow\right\rangle\left\vert A_{\nwarrow}\right\rangle +\beta\left\vert \nearrow\right\rangle\left\vert A_{\nearrow}\right\rangle $, where the linearity of quantum mechanics was taken into account (we note that the states of the apparatus $\left\vert A_{\nwarrow}\right\rangle $ and $\left\vert A_{\nearrow}\right\rangle $ are not necessarily orthogonal). The state of $\mathcal{A}$ now contains a record of $\mathcal{S}$, indicated by the subindexes $\nwarrow$ and $\nearrow$. Using the fact that the norm of the state of the composite system must be preserved (and recognizing that $\left\langle A_{blank} |A_{blank}\right\rangle =\left\langle A_{\nwarrow}|A_{\nwarrow}\right\rangle =\left\langle A_{\nearrow}|A_{\nearrow}\right\rangle =1$), the relation $\left\langle \psi_{\mathcal{S}}|\psi_{\mathcal{S}}\right\rangle -\left\langle \Phi_{\mathcal{SA}}|\Phi_{\mathcal{SA}}\right\rangle =0$ yields
\begin{equation}
\left\langle \nwarrow|\nearrow\right\rangle \left(  1-\langle A_{\nwarrow}|A_{\nearrow}\rangle\right)  =0.
\label{eq:outcome-states-orthogonality}
\end{equation}
From condition (\ref{eq:outcome-states-orthogonality}), it is possible to conclude that if $\left\langle \nwarrow|\nearrow\right\rangle \neq0$ the measurement apparatus can carry no imprint of the system state, since $\langle A_{\nwarrow}|A_{\nearrow}\rangle=1$ must hold. On the other hand if $\left\langle \nwarrow|\nearrow\right\rangle =0$, this allows the detector to carry imprints of the system state (since $\langle A_{\nwarrow}|A_{\nearrow}\rangle$ can be arbitrary now, including the prefect record case where $\langle A_{\nwarrow}|A_{\nearrow}\rangle=0$). This result is independent of the purity of the initial apparatus state \cite{Zurek}.

In order to study the effect of the choice of reference frame in the measurement process, we consider what happens from the particle's point of view. We regard the measurement apparatus as a semiclassical detector with the simplest possible two-state structure starting in a blank file state, which for a static observer, may be represented by $\int d\mathbf{p}\,a_{blank}\left(  \mathbf{p}\right)  |A_{blank},\mathbf{p}\rangle$. The measurement process is\ now described through the relations $\left(  \alpha\left\vert\nwarrow\right\rangle +\beta\left\vert \nearrow\right\rangle \right)  \int d\mathbf{p}\,a_{\text{\emph{blank}}}\left(  \mathbf{p}\right)  \left\vert A_{\text{\emph{blank}}},\mathbf{p}\right\rangle \Longrightarrow\left\vert \Phi_{\mathcal{SA}}\right\rangle =\alpha\left\vert \nwarrow\right\rangle \int d\mathbf{p}\,a_{\nwarrow}\left(  \mathbf{p}\right)  \left\vert A_{\nwarrow},\mathbf{p}\right\rangle +\beta\left\vert \nearrow\right\rangle \int d\mathbf{p}\,a_{\nearrow}\left(  \mathbf{p}\right)  \left\vert A_{\nearrow},\mathbf{p}\right\rangle $, where the apparatus has a very narrow Gaussian momentum distribution, i.e., $a_{\emph{blank}}\left(  \mathbf{p}\right)  =\sqrt{2} a_{\nwarrow}\left(  \mathbf{p}\right)  =\sqrt{2} a_{\nearrow}\left(  \mathbf{p}\right)  =a\left(\mathbf{p}\right)  $ given by Eq. (\ref{momentum-dist}). Let us suppose, in the static scenario, that the detector is ideal and performs a perfect measurement, in other words, Tr$\rho_{\mathcal{A}_{\nwarrow}}\rho_{\mathcal{A}_{\nearrow}}=0$, where $\rho_{\mathcal{A}_{\lambda}}=\int d\mathbf{p\Phi}_{\lambda}\left(  \mathbf{p}\right)  \mathbf{\Phi}_{\lambda}^{\ast}\left(  \mathbf{p}\right)  $ with $\mathbf{\Phi}_{\mathbf{\lambda}}\left(  \mathbf{p}\right)  \equiv\int d\mathbf{p}\,a_{\lambda}\left(\mathbf{p}\right)  \left\langle \mathbf{p}\left\vert A_{\lambda},\mathbf{p}\right.  \right\rangle $ ($\lambda=\nwarrow,\nearrow$),  and $|\nwarrow\rangle$ is in the direction of the motion and $|\nearrow\rangle$ is orthogonal to $|\nwarrow\rangle$.

Regarding a moving observer relative to the qubit's source, an interesting feature occurs from the viewpoint of the qubit's reference frame. From such a point of view, the quantum degrees of freedom of the detector apparatus will suffer a Wigner rotation in such a way that the transformed composite state is
\begin{align}
\left\vert \widetilde{\Phi}_{\mathcal{SA}}\right\rangle  & =\alpha\left\vert\nwarrow\right\rangle \int d\mathbf{p}\,b_{\nwarrow}\left(  \mathbf{p}\right)\left\vert A_{\nwarrow},\mathbf{p}\right\rangle \nonumber\\
& +\beta\left\vert \nearrow\right\rangle \int d\mathbf{p}\,b_{\nearrow}\left(\mathbf{p}\right)  \left\vert A_{\nearrow},\mathbf{p}\right\rangle,
\end{align}
with $b_{\lambda}(\mathbf{p})$ given in Eq. (\ref{eq:spinor-component-transformation}). Once more, using the fact that the norm of the state of the composite system must be preserved and tracing out the momentum degrees of freedom, the relation $\left\langle \psi_{\mathcal{S}}|\psi_{\mathcal{S}}\right\rangle -\left\langle \widetilde{\Phi}_{\mathcal{SA}}|\widetilde{\Phi}_{\mathcal{SA}}\right\rangle =0$ gives us
\begin{equation}
\left\vert \left\langle \nwarrow|\nearrow\right\rangle \right\vert ^{2}\left(\text{Tr}\widetilde{\rho}_{\mathcal{A}_{blank}}^{2}-\text{Tr}\widetilde{\rho}_{\mathcal{A}_{\nwarrow}}\widetilde{\rho}_{\mathcal{A}_{\nearrow}}\right)=0,
\label{eq:mixed-states-orthogonality}
\end{equation}
with $\widetilde{\rho}_{\mathcal{A}_{\lambda}}=\int d\mathbf{p}\widetilde{\mathbf{\Phi}}_{\lambda}\left(  \mathbf{p}\right)  \widetilde{\mathbf{\Phi}}_{\lambda}^{\ast}\left(  \mathbf{p}\right)  $,  $\widetilde{\mathbf{\Phi}}_{\lambda}\left(  \mathbf{p}\right)  \equiv\int d\mathbf{p}\,b_{\lambda}\left(  \mathbf{p}\right)  \left\langle \mathbf{p}\left\vert A_{\lambda},\mathbf{p}\right.  \right\rangle $. In Eq. (\ref{eq:mixed-states-orthogonality}) $\widetilde{\rho}_{\mathcal{A}_{blank}}$ is the transformed apparatus reduced density operator in the blank state and
\begin{equation}
\widetilde{\rho}_{\mathcal{A}_{\nwarrow}}=\frac{1}{2}\left(
\begin{array}
[c]{cc}
1+\widetilde{\gamma} & \widetilde{\delta}\\
\widetilde{\delta}^{\ast} & 1-\widetilde{\gamma}
\end{array}
\right),
\end{equation}
and
\begin{equation}
\widetilde{\rho}_{\mathcal{A}_{\nearrow}}=\frac{1}{2}\left(
\begin{array}
[c]{cc}
1+\widetilde{\gamma} & -\widetilde{\delta}\\
-\widetilde{\delta}^{\ast} & 1-\widetilde{\gamma}
\end{array}
\right),
\end{equation}
are the post-measurement apparatus reduced states, with $\widetilde{\gamma}\equiv\int d\mathbf{p}\left(\left\vert b_{\nwarrow}(\mathbf{p})\right\vert ^{2}-\left\vert b_{\nearrow}(\mathbf{p})\right\vert ^{2}\right)  $ and $\widetilde{\delta}\equiv2\int d\mathbf{p}b_{\nwarrow}(\mathbf{p})b_{\nearrow}(\mathbf{p})^{\ast}.$ Therefore, from the point of view of the particle, the degrees of freedom of the moving detector are in a mixed state. 

In the nonstatic scenario, we can draw, from condition (\ref{eq:mixed-states-orthogonality}), conclusions very similar to those obtained in the previous static scenario [Eq. (\ref{eq:outcome-states-orthogonality})], since either $\left\langle \nwarrow|\nearrow\right\rangle =0$ (orthogonality of outcomes) or $\widetilde{\rho}_{\mathcal{A}_{\nwarrow}}=\widetilde{\rho}_{\mathcal{A}_{\nearrow}}$ (no imprint in the apparatus) is necessary to Eq. (\ref{eq:mixed-states-orthogonality}) to hold \cite{Zurek}. Thus, from the point of view of the particle's reference frame, due to the Wigner rotation suffered by the detector quantum degrees of freedom, the measurement apparatus will behave as if it was not ideal, loosing information in the measurement processes. In this case, the probability of detecting the spin states $\left\vert \nwarrow\right\rangle $ and $\left\vert \nearrow\right\rangle $ by the moving observer (with velocity $v$ parallel to the spin orientation) will be given respectively by $\widetilde{\text{}\mathcal{P}}_{\nwarrow}=\eta\mathcal{P}_{\nwarrow}$ and $\widetilde{\mathcal{P}}_{\nearrow}=(1-\eta)\mathcal{P}_{\nearrow},$ where $\mathcal{P}_{\nwarrow}$ and $\mathcal{P}_{\nearrow}$ are the respective probabilities for the static observer to detect $\left\vert \nwarrow\right\rangle $ and $\left\vert \nearrow\right\rangle $ and $\eta=\left\vert K\left(  \left.  q^{0}\right/p^{0}\right)  ^{1/2}\left[  \left(  q^{0}+m\right)  \cosh\left(  \left. \theta\right/  2\right)  +q^{x}\sinh\left(  \left.  \theta\right/  2\right) \right]  \right\vert ^{2}$ is the quantum efficiency of the moving detector. Although both observers are equipped with perfect detectors (in the static scenario), when one of them is in relative motion with respect to the source $\mathbb{S}$, the effective measurement processes occurs with a nonideal efficiency $\eta<1$. 

\section{Discussions} 
The lack of information perceived by a moving observer is due to some correlation between the spin and momentum of the particle induced by the Wigner rotation. Such an effect is expected to occur in every system with at least one degree of freedom besides the momentum. These observations may be extended for higher dimension systems. As previously observed, the amount of information acquired in the measurement processes depends on the reference frame of the observer (detector). The moving detector has an \emph{effective efficiency} due to a net effect of the Wigner rotation. However, all observers will agree about the orthogonal character of the possible outcomes in measurement processes as depicted in condition (\ref{eq:mixed-states-orthogonality}). This feature of the wave function collapse is Lorentz invariant. These conclusions could have important consequences for the development of quantum communication devices involving moving partners. Concerning fundamental aspects of quantum mechanics, this observation implies that, independently of the observer, we must establish some orthogonal basis in order to define measurement outcomes.

Finally, we expect that the decoherence process that affects open quantum systems does not modify the conclusions of this paper, since we can always describe the whole system as a closed one, by including in it the reservoir degrees of freedom. We also mention that the analysis of the case of accelerated observers could instigate interesting questions. This case may present novel features due to the well known Unruh effect \cite{Matsas} which leads to a degradation of the correlations presented in composed quantum systems \cite{Landulfo,Lucas}.

\section*{Acknowledgements}
RMS and LCC thank V. Vedral for stimulating discussions. We also thank A. G. Dias, A. G. S. Ladulfo and G. E. A. Matsas for a critical reading of this manuscript. We are grateful for the financial support from UFABC, CAPES, and FAPESP. This work was performed as part of the Brazilian National Institute of Science and Technology for Quantum Information (INCT-IQ).

\end{document}